# Angular constraint on light-trapping absorption enhancement in solar cells

Zongfu Yu and Shanhui Fan

Ginzton Lab, Stanford University, Stanford, CA, 94305

Abstract:

Light trapping for solar cells can reduce production cost and improve energy conversion efficiency. Understanding some of the basic theoretical constraints on light trapping is therefore of fundamental importance. Here, we develop a general angular constraint on the absorption enhancement in light trapping. We show that there is an upper limit for the angular integration of absorption enhancement factors. This limit is determined by the number of accessible resonances supported by an absorber.

Keywords: photovoltaics; light trapping; nanostructure solar cells; absorption enhancement;

Light trapping enhances the absorption of active materials in photovoltaic cells. The use of light trapping results in a thinner active region in a solar cell, which lowers the production cost by reducing the amount of material used, and increases the energy conversion efficiency by facilitating carrier collection and enhancing the open circuit voltage [1]. Understanding various theoretical limits on absorption enhancement is therefore of fundamental importance in solar cell design.

In this Letter, we provide a general theoretical constraint on the angular dependency of light trapping enhancement factors. Understanding the angular dependency of light trapping performance is important in practice: sun light has a significant diffusive component; roof-top system typically has only limited tracking capabilities; and high-concentration system requires operation in a range of angles. While an upper limit of absorption enhancement has been obtained, the classical light trapping theory [2-3] assumed an angle-independent absorption coefficient within an absorption cone, and is therefore difficult to apply to practical solar cells that have a more complex angular response. Here, using the statistical coupled-mode theory developed in [4], we derive a general limit for arbitrary angular response. This upper limit applies to both random textured and periodic absorbers.

To model light trapping in solar cells, we consider an absorbing film with air on one side and a strongly reflecting mirror placed on the other. The film has a refractive index of $n$, a material absorption constant $\alpha$, and a thickness $d$. Following [2-4], we consider the weakly absorbing case where $\alpha d \ll 1$, i.e. the single pass absorption is negligible. Most of the results in this paper also assume that the thickness $d \gg \lambda$.

In this system, light trapping is accomplished with the use of either random roughness[2-3] or periodic grating [4-13] on either the top or the bottom surfaces of the film (Fig. 1a,b) The performance of the light trapping is quantified by the absorption enhancement factor $f = \frac{A}{\alpha d}$, where $A$ is the absorption coefficient in the presence of the light trapping scheme. The standard light trapping theory shows that the maximum absorption enhancement factor is $f_{\max} = 4n^2 / \sin^2(\theta_c)$. Here $\theta$ is half of apex angle of the absorption cone. In deriving this limit, one assumes that light with an incident direction within the absorption cone has absorption that is independent of angle of incidence, while for light coming outside the absorption cone, there is no absorption. $f_{\max}$ reduces to $4n^2$ for the isotropic case [2] where $\theta_c = 90^0$.

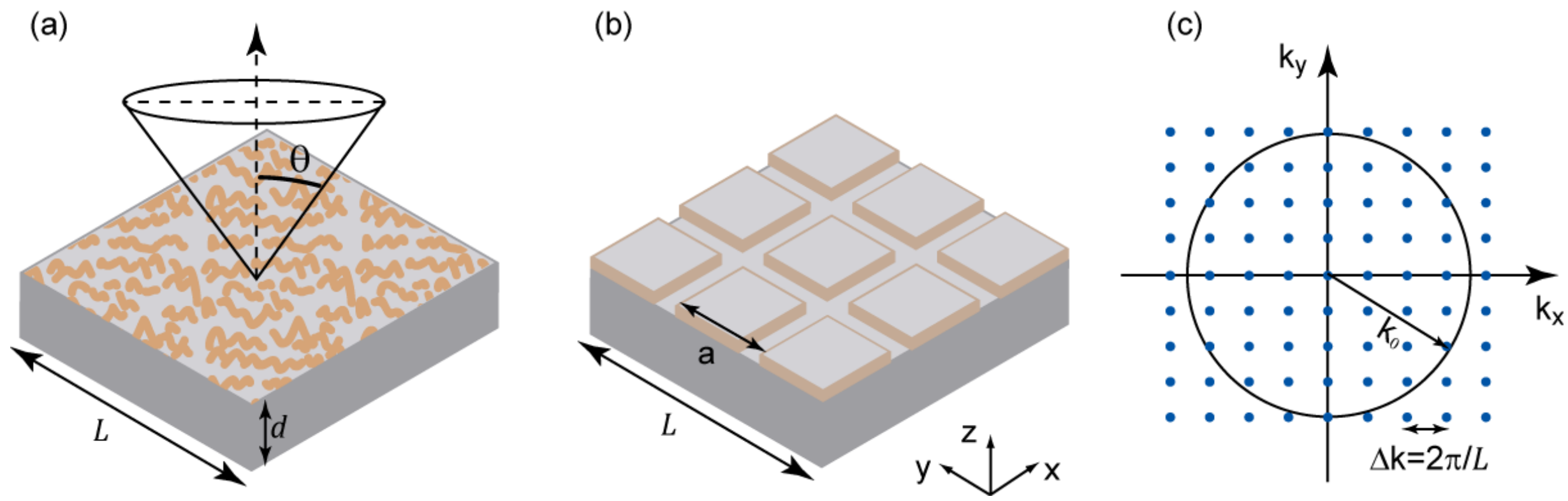

Fig. 1. Schematic of light trapping with (a) random textures, and (b) periodic grating. (c) Free space channels in $k_x$-$k_y$ space

The standard light-trapping theory thus describes a tradeoff between the absorption enhancement and the angular response of the absorber: the narrower the angular response range, the higher the potential absorption enhancement. The derivation in [3] however is somewhat restrictive: it is derived assuming an isotropic angular response within the absorption cone. Practical geometries, on the other hand, can have more complex angular response[13-14]. Moreover, the derivation in [3] is not applicable when the grating periodicity is comparable to the wavelength, which has been the subject of a large number of recent studies[8, 11-13]. In this letter, we overcome these limitations of the standard theory, by showing a general angular constraint for light trapping enhancement:

$$\mathrm{F} \equiv \int_0^{\pi/2} d\theta \int_0^{2\pi} d\varphi f(\theta,\varphi)\cos(\theta)\sin(\theta) \le 4\pi n^2 \tag{1}$$

where $f(\theta,\varphi)$ is the absorption enhancement factor, averaged over the two polarizations, for the light incident from a direction determined by the incident angle $\theta$ and the azimuthal angle $\varphi$. The general relation in Eq. (1) applies to all light trapping schemes independent of their angular response.

We use the theoretical frame work recently developed in [4] to prove above constraint. Light absorption is treated by considering the interaction between the optical resonances in an absorber and incident waves from free space. We start with a general situation where the absorber has a large side length $L$ (Fig. 1a, b). To facilitate the analytic calculations, as is standard, we assume a periodic boundary condition with a periodicity $L$. The absorber supports a total number $\rho(\omega)\Delta\omega$ of optical resonances in the frequency range $[\omega, \omega+\Delta\omega]$. When calculating the density of states $\rho(\omega)$, it is sufficient to approximate these resonant modes as plane waves in a bulk medium with refractive index $n$, which results in:

$$\rho(\omega) = 2\frac{4\pi n^3\omega^2}{c^3}\left(\frac{L}{2\pi}\right)^2\left(\frac{d}{2\pi}\right) \tag{2}$$

where $c$ is the light speed in vacuum.

The free space has $N$ distinct channels that carry the incident and outgoing plane waves. Each channel is characterized by a parallel wavevector $\vec{k}_{//}$ specifying the incidence direction. With our assumption of a periodic boundary condition, these wavevectors form a lattice in the $k_x - k_y$ space with a lattice constant $\Delta k = 2\pi / L$ (Fig. 1c). In addition, $\left|\vec{k}_{//}\right| < \omega / c$ since these waves are propagating waves in free space. For each parallel wavevectors, there are two orthogonal polarizations, and hence two independent channels.

We first consider a plane wave incident upon the structure from $n$-th channel. The resulting amplitude $a_m$ in the $m$-th resonance is then described by the coupled-mode theory equation: [15]

$$\frac{d}{dt}a_m = (j\omega_m - \frac{\sum_{s=1}^{N}\gamma_{m,s} + \gamma_0}{2})a_m + j\sqrt{\gamma_{m,n}}S_n \tag{3}$$

Here the resonance amplitude is normalized such that $|a_m|^2$ is the energy per unit area in the film. $\omega_m$ is the resonant frequency. $S_n$ is the amplitude of the incident plane wave, with $|S_n|^2$ corresponding to its intensity. $\gamma_0 = \alpha_0 \frac{c}{n}$ is the intrinsic absorption rate of the resonances determined only by the material used in the absorber. In contrast to the isotropic absorption cone assumption made in [2, 4], here we do not make any assumption about $\gamma_{m,n}$ -- in general they can be all different.

From Eq.(3), the absorption by the $m$-th resonance when light is incident from the $n$-th channel is

$$A_{m,n}(\omega) = \gamma_0 \left|\frac{a_m(\omega)}{S_n(\omega)}\right|^2 = \frac{\gamma_{m,n}\gamma_0}{(\omega-\omega_0)^2 + \left(\sum_{s=1}^{N}\gamma_{m,s} + \gamma_0\right)^2 / 4} \tag{4}$$

The *spectral cross section* [4], which characterizes the contribution of a resonance to broadband absorption enhancement, is calculated as

$$\sigma_{m,n} = \int_{-\infty}^{\infty} A_{m,n}(\omega)d\omega = 2\pi \frac{\gamma_{m,n}\gamma_0}{\sum_{s=1}^{N}\gamma_{m,s} + \gamma_0} \tag{5}$$

Summing over the contributions of all resonances in the frequency range of $[\omega, \omega + \Delta\omega]$, the light trapping enhancement factor for light incident from the $n$-th channel is therefore:

$$f_n = \sum_m \frac{\sigma_{m,n} / \Delta\omega}{\alpha_0 d} \tag{6}$$

Further summing over all channels, and using Eq. (5), we obtain the following relation:

$$\sum_n f_n = \frac{2\pi\gamma_0}{\alpha_0 d\Delta\omega}\sum_m \frac{\sum_{s=1}^{N}\gamma_{m,s}}{\sum_{s=1}^{N}\gamma_{m,s}+\gamma_0} \le \frac{2\pi c}{nd}\rho(\omega) \tag{7}$$

The equal sign is obtained when $\sum_{s=1}^{N}\gamma_{m,s} >> \gamma_0$ for any $m$, i.e. all the resonances are in the over-coupling regime. Eq.(7) shows that the summation of the light trapping enhancement factors over all channels is limited by the total density of state $\rho(\omega)$.

Now we convert the k-space summation in Eq. (7) to an angular integration. Using the relation

$$\begin{aligned} k_x &= k_0 \sin(\theta)\cos(\varphi) \\ k_y &= k_0 \sin(\theta)\sin(\varphi) \end{aligned} \tag{8}$$

where $k_0 = \omega / c$, we can rewrite angular integration for $\mathrm{F}$ in Eq.(1) as

$$\mathrm{F} = \iint_{k_x^2+k_y^2\le k_0^2} \frac{1}{k_0^2} f(k_x,k_y)dk_x dk_y = \frac{1}{2}\frac{\Delta k^2}{k_0^2}\sum_n f_n \tag{9}$$

The factor 1/2 comes from the average of two polarizations. Combining Eq.(7) and (9), we obtain

$$\mathrm{F} \le \frac{c}{2k_0^2 n}\left(\frac{2\pi}{L}\right)^2 \frac{2\pi}{d}\rho(\omega) \tag{10}$$

Substituting Eq. (2) into Eq. (10), we then obtain $\mathrm{F} \le 4\pi n^2$ and thus prove Eq.(1).

We emphasize that Eq. (1) applies to both random and periodic textures. In the case of random texturing, light incident from a given channel $n$ typically has access to all resonances, i.e. $\gamma_{m,n} \ne 0$ for all $m$. Hence all resonances, in principle, can contribute to the summation over $m$ in Eqs. (6) and (7). The upper limit in Eq. (1) is then reached, when all resonances are in the over-coupled regime. We note that this result does not assume a particular angular response function.

In the case of periodic texture, due to the translational symmetry in periodic structures, some of the resonances may not be accessible to free space, leading to a reduced upper limit of the angular integration $\mathrm{F}$. Suppose that light is incident from channel $n$ with a parallel wavevector $\vec{k}_n$ satisfying $\left|\vec{k}_n\right| \le k_0$, such incident light can only couple to resonances with their parallel wavevectors satisfying

$$\vec{k}_m = \vec{k}_n + \vec{G} \tag{11}$$

where $G$ represents all reciprocal lattice vectors. Those resonances that do not satisfy Eq. (11) for every $n$ do not contribute to light trapping.

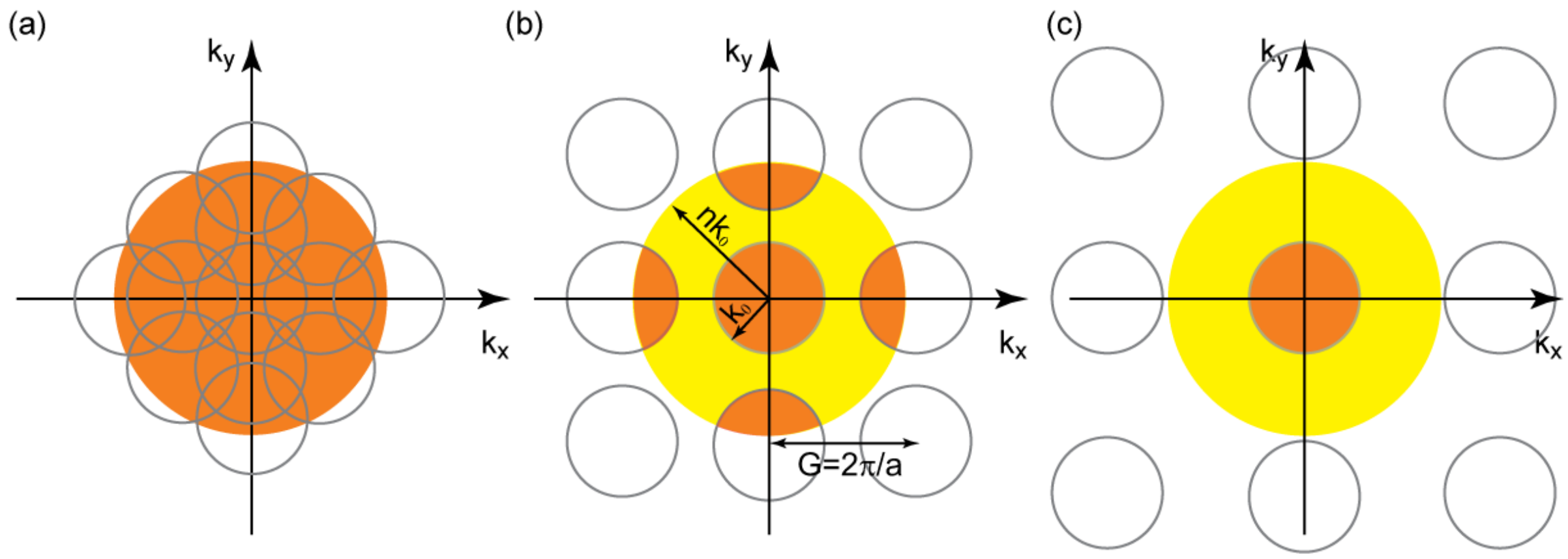

Fig. 2. Schematic of contributing resonances for different periodicities. The contributing resonances are determined by the overlap region (orange regions) between the resonance region (yellow regions) and the radiation zone (gray circles) (a) $a >> \lambda$. (b) $a \sim \lambda$. (c) $a << \lambda$.

We now illustrate the constraint of Eq. (11) graphically, by considering a concrete example of a film textured with a square lattice grating having a periodicity *a*. (Fig. 1b). At a given frequency ω, the plane waves in the film have their parallel wavevectors filling a solid circle with a radius of $nk_0$ in the $k_x - k_y$ space (yellow regions in Fig. 2). We refer to this solid circle as the *resonance region*. Since the absorber is periodic in-plane, the parallel wavevector of the resonances is rigorously defined by the Bloch theorem. The plot here represents the Bloch wavevector in the extended zone scheme. On the other hand, the free-space channels, in the $k_x - k_y$ space, cover a periodic array of filled circles, each having a radius of $k_0$, and centered at one of the reciprocal lattice point (gray circles in Fig. 2). We refer to these arrays of circles as the *radiation zone*. Only those resonances with their parallel wavevectors falling into the radiation zone can couple to free space (Orange regions in Fig. 2).

At a given wavelength λ, we first consider the case of a large period, $a >> \lambda$. Examining Fig. 2a, we note that all resonances can contribute to light trapping, and this leads to the maximum $\mathrm{F} = 4\pi n^2$ . At a smaller period $a \le \lambda\sqrt{2}/2$, some resonances fall outside the radiation zone (Fig. 2b), which do not contribute to light trapping. Therefore, the overall angle-integrated enhancement factor F is lower. In general, as the period *a* decreases, the number of total contributing resonances and thus F decrease. When $a < \lambda/(n+1)$, the resonance region has no overlap with the radiation zone except for the circle centered at $\vec{k}_{//} = 0$ (Fig. 2c). In this regime, the upper limit of F remains a constant independent of the periodicity *a*.

The numerically calculated upper limit F, as obtained by a direct evaluation of Eq. (9) and shown in Fig. 3, indeed follows the general trend as graphically illustrated in Fig. 2. With a square lattice periodicity, F drops below $4\pi n^2$ when $a/\lambda \le \sqrt{2}/2$. At small periods, angular integration decreases to a constant $\mathrm{F} = 4\pi n^2(1-\sqrt{1-1/n^2})$ that is lower at higher *n*. Similar trend is observed for other periodic lattices.

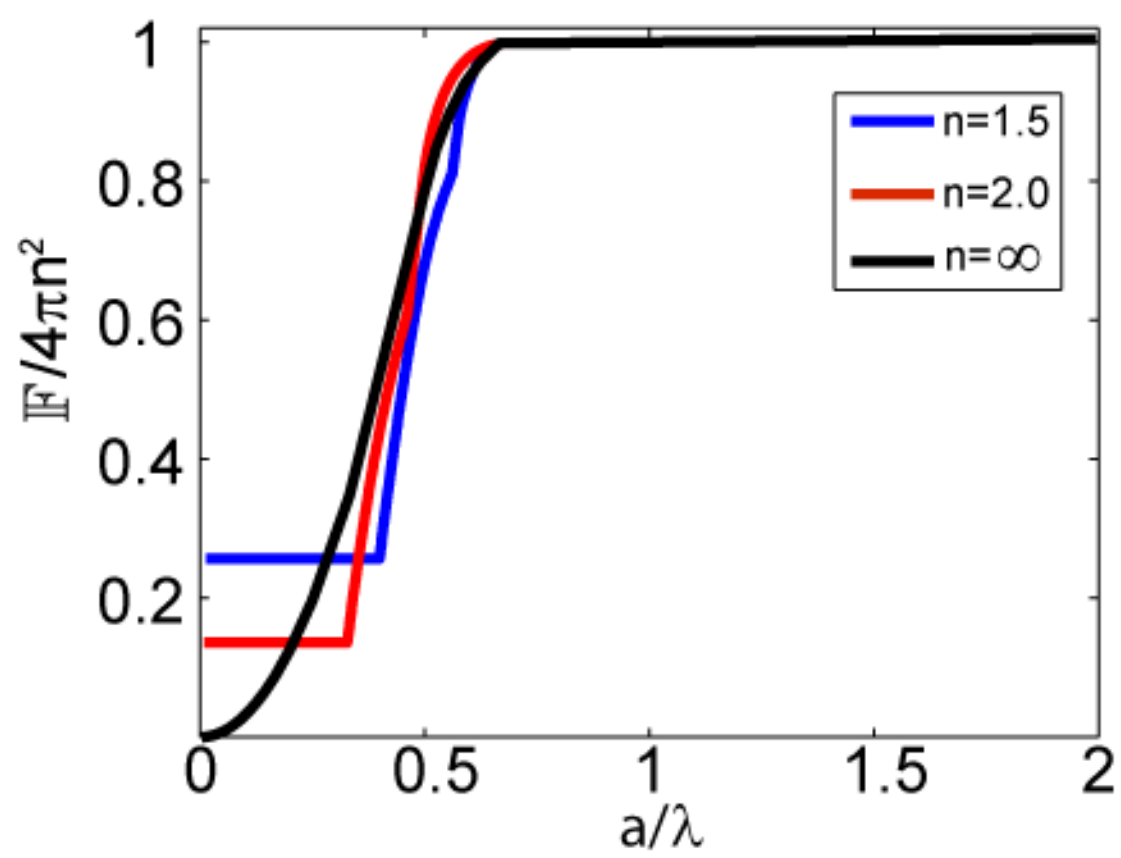

Fig. 3. Limit of angular integration of enhancement factor for square lattice.

As a specific application of Eq. (1), we calculate the upper limit of light trapping in a diffusive sunlight condition. The isotropic diffusive radiation is described by the intensity profile: $I(\theta,\varphi)\mathrm{d}\Omega = I_0\mathrm{d}\Omega$, where $\mathrm{d}\Omega = \sin(\theta)\mathrm{d}\theta\mathrm{d}\varphi$ is the solid angle, and $I_0$ is the intensity of the radiation per unit solid angle. We consider a film of absorber with thickness *d*. The total absorption is:

$$A_t = \int_{hemisphere} A(\theta,\varphi) I_0 \cos(\theta) d\Omega \tag{12}$$

where $A(\theta,\varphi) = f(\theta,\varphi)\alpha_0 d$ is absorption coefficients. $I_0\cos(\theta)$ is the light intensity incident upon the surface of the absorber. Using Eq.(1), we obtain the total absorption as

$$A_t \le 4n^2\pi d\alpha_0 I_0 \tag{13}$$

The upper limit of total light absorption in isotropic diffusive radiation is independent of the angular response of the absorber. This conclusion also applies to the radiation profile with full solar concentration.

We conclude by making a few remarks. The existence of non-radiative modes in a film with sub-wavelength periodic texturing is quite well known [16]. Here, we point out that having such non-radiative modes in a solar cell lowers the light trapping enhancement factor, by reducing the total number of resonances that are accessible to external radiation. The results here further show that if there is a need to operate over broad angular range, one should not use grating structures with periodicity less than a wavelength. Finally, while Eq.(1) is applicable for a film that is internally uniform and has a thickness of a few wavelengths, Eq.(7), involving the density of states, is completely general and applies to any light trapping schemes. Thus, using the nanophotonic light trapping schemes as predicted in [4, 17], one could achieve angular integrated enhancement factor significantly beyond $4\pi n^2$. In structures such as nanowires [18-24], where the density of states is significantly different from uniform mediums, the enhancement factor $\mathrm{F}$ should also be different from $4\pi n^2$.

This work was supported by DOE Grant No. DE- FG02-07ER46426